\documentclass[journal]{IEEEtran}
\usepackage[english]{babel}

\usepackage{ifpdf}

\usepackage{cite} % Orders citations.
\usepackage{url}
\usepackage{hyperref}

\ifCLASSINFOpdf
	\usepackage[pdftex]{graphicx}
	\graphicspath{{./figures/}}
\else
	\usepackage[dvips]{graphicx}
	\graphicspath{./figures/}
\fi
\usepackage{color}
\usepackage{pgf, tikz, pgfplots}
\usetikzlibrary{shapes, arrows, automata}
\usetikzlibrary{calc,hobby,decorations}
\usepackage[cmex10]{amsmath}
\usepackage{amsfonts, amssymb, amsthm}
\usepackage{mathrsfs}
\usepackage{mathbbol}
\usepackage{adjustbox}

\usepackage{array}
\usepackage{enumerate}
\usepackage{multirow}
\usepackage{rotating}
\usepackage{subcaption}
\usepackage{dsfont}
\usepackage{needspace}

\input{mysymbol.sty}

\definecolor{penndarkestblue}{cmyk}{1,0.74,0,0.77}
\definecolor{penndarkerblue}{cmyk}{1,0.74,0,0.70}
\definecolor{pennblue}{cmyk}{0.99,0.66,0,0.57} 
\definecolor{pennlighterblue}{cmyk}{0.98,0.44,0,0.35}
\definecolor{pennlightestblue}{cmyk}{0.38,0.17,0,0.17} 
\definecolor{penndarkestred}{cmyk}{0,1,0.89,0.66}
\definecolor{penndarkerred}{cmyk}{0,1,0.88,0.55}
\definecolor{pennred}{cmyk}{0,1,0.83,0.42} 
\definecolor{pennlighterred}{cmyk}{0,1,0.6,0.24}
\definecolor{pennlightestred}{cmyk}{0,0.43,0.26,0.12} 
\definecolor{penndarkestgreen}{cmyk}{1,0,1,0.68}
\definecolor{penndarkergreen}{cmyk}{1,0,1,0.57}
\definecolor{penngreen}{cmyk}{1,0,1,0.44} 
\definecolor{pennlightergreen}{cmyk}{1,0,1,0.25}
\definecolor{pennlightestgreen}{cmyk}{0.43,0,0.43,0.13}
\definecolor{penndarkestorange}{cmyk}{0,0.65,1,0.49}
\definecolor{penndarkerorange}{cmyk}{0,0.65,1,0.33}
\definecolor{pennorange}{cmyk}{0,0.54,1,0.24} 
\definecolor{pennlighterorange}{cmyk}{0,0.32,1,0.13}
\definecolor{pennlightestorange}{cmyk}{0,0.15,0.46,0.06}
\definecolor{penndarkestpurple}{cmyk}{0,1,0.11,0.86}
\definecolor{penndarkerpurple}{cmyk}{0,1,0.13,0.82}
\definecolor{pennpurple}{cmyk}{0,1,0.11,0.71} 
\definecolor{pennlighterpurple}{cmyk}{0,1,0.05,0.46}
\definecolor{pennlightestpurple}{cmyk}{0,0.35,0.02,0.23}
\definecolor{pennyellow}{cmyk}{0,0.20,1,0.05} 
\definecolor{pennlightgray1}{cmyk}{0,0,0,0.05}
\definecolor{pennlightgray3}{cmyk}{0.01,0.01,0,0.18}
\definecolor{pennmediumgray1}{cmyk}{0.04,0.03,0,0.31}
\definecolor{pennmediumgray4}{cmyk}{0.08,0.06,0,0.54}
\definecolor{penndarkgray2}{cmyk}{0.09,0.07,0,0.71}
\definecolor{penndarkgray4}{cmyk}{0.1,0.1,0,0.92}
\newcommand{\vertiii}[1]{{\left\vert\kern-0.25ex\left\vert\kern-0.25ex\left\vert #1 \right\vert\kern-0.25ex\right\vert\kern-0.25ex\right\vert}}

\newtheorem{lemma}{\hspace{0pt}\bf Lemma}
\newtheorem{proposition}{\hspace{0pt}\bf Proposition}

\newtheorem{remark}{\hspace{0pt}\bf Remark}

\newtheorem{innercustomgeneric}{\customgenericname}
\providecommand{\customgenericname}{}
\newcommand{\newcustomtheorem}[2]{%
  \newenvironment{#1}[1]
  {%
   \renewcommand\customgenericname{#2}%
   \renewcommand\theinnercustomgeneric{##1}%
   \innercustomgeneric
  }
  {\endinnercustomgeneric}
}

\newcustomtheorem{customthm}{Theorem}
\newcustomtheorem{customlemma}{Lemma}

\newcommand{\lfoc}[1]{{\color{blue}\textbf{LFOC:} #1}}

\begin{document}

\title{Reply to `Comments on Graphon Signal Processing'}

\author{Luana~Ruiz, Luiz~F.~O.~Chamon~%\IEEEmembership{Student~Member,~IEEE,}}
        and~Alejandro~Ribeiro%,~\IEEEmembership{Member,~IEEE}% <-this % stops a space
%\thanks{ L. Ruiz and A. Ribeiro are with the Dept. of Electrical and Systems Eng., Univ. of Pennsylvania. L. F. O. Chamon is with the Simons Institute, U. C. Berkeley.  %Email: \{rubruiz,luizf,aribeiro\}@seas.upenn.edu.
%}
}

% Headers:
%\markboth{IEEE TRANSACTIONS ON SIGNAL PROCESSING (ACCEPTED)}%
%{Transferability}

\maketitle

%\begin{abstract}
%Graph neural networks (GNNs) are composed of layers consisting of graph convolutions and pointwise nonlinearities. Due to their invariance and stability properties, GNNs are provably successful at learning representations from data supported on moderate-scale graphs. However, they are difficult to learn on large-scale graphs. In this paper, we study the problem of training GNNs on graphs of moderate size and transferring them to large-scale graphs. We use graph limits called graphons to define limit objects for graph filters and GNNs---graphon filters and graphon neural networks ($\bbW$NNs)---which we interpret as generative models for graph filters and GNNs. We then show that graphon filters and $\bbW$NNs can be approximated by graph filters and GNNs sampled from them on weighted and stochastic graphs. Because the error of these approximations can be upper bounded, by a triangle inequality argument we can further bound the error of transferring a graph filter or a GNN across graphs. 
%Our results show that (i) the transference error decreases with the graph size, and (ii) that graph filters have a transferability-discriminability tradeoff that in GNNs is alleviated by the scattering behavior of the nonlinearity. These findings are demonstrated empirically in a recommendation problem and in a decentralized control task.
%\end{abstract}

%\begin{IEEEkeywords}
%graph neural networks, transferability, graph signal processing, graphons
%\end{IEEEkeywords}

\IEEEpeerreviewmaketitle

%%%%%%%%%%%%%%%%%%%%%%%%%%%%%%%
%%% SECTION : Introduction  %%%
%%%%%%%%%%%%%%%%%%%%%%%%%%%%%%%

\section{Introduction} \label{sec:intro}

This technical note addresses an issue with the proof (but not the statement) of \cite[Proposition 4]{ruiz2020graphonsp}. The statement of the proposition is correct, but the proof as written in \cite{ruiz2020graphonsp} is not and due to a typo in the manuscript, a reference to the correct proof is effectively missing. In the sequel, we present \cite[Proposition 4]{ruiz2020graphonsp} and its proof. The proof follows from results in \cite{jansen} that we reproduce here for clarity of exposition. 

Since the statement of the proposition remains correct, no change in the results of \cite{ruiz2020graphonsp} are required. In particular, Lemma 3 and Lemma 4 showing spectral convergence of graphs to graphons, Theorem 1 showing convergence of the GFT to the WFT, and Theorems 3 and 4 showing convergence of graph to graphon filters, remain valid.

\section{Proposition 4 of \cite{ruiz2020graphonsp}}\label{sec_proposition}

A graphon is a bounded symmetric measurable function $\bbW:[0,1]^2\to[0,1]$ which we interpret as a graph with an uncountable label set. For a graphon $\bbW$, we define the cut norm as per~\cite[Equation (7)]{ruiz2020graphonsp} and~\cite[Equation (4.2)]{jansen}
\begin{align}\label{eqn_cut_norm}
   \|\bbW \|_{\square,1} 
      := \sup_{S,T \subseteq [0,1]}
           \left | 
               \int_{S \times T} \bbW(u,v) \, du dv 
                  \right |.
\end{align}
Note that~\cite{ruiz2020graphonsp} uses~$\|{\bbW}\|_{\square} = \|{\bbW}\|_{\square,1}$ to denote the cut norm. We change the notation here because we will need to introduce other cut norm versions. Associated to the graphon~$\bbW$, we define an integral linear operator $T_\bbW: L^2 \to L^2$ whose action on a function~$X \in L^2$ is defined as
\begin{equation}\label{eqn_operator}
    T_\bbW X = \int_{0}^{1} \bbW(u,v) X(u) du
        \text{.}
\end{equation}
Its operator norm is then given by
\begin{align}\label{eqn_22norm}
   \vertiii{{T_\bbW}}_{2,2}
      := \sup_{\| X \|_{L^2} \leq 1} 
            \left\| 
               \int_{0}^{1} \bbW(u,v) X(u) \, du 
                  \right\|_2.
\end{align}
Once again, note that $\vertiii{{T_\bbW}}_{2,2} = \vertiii{{T_\bbW}}$ in~\cite{ruiz2020graphonsp}. In the above notation, Proposition 4 in \cite{ruiz2020graphonsp} claims the following:

%%%%%%%%%%%%%%%%%%%%%%%%%%%%%%%%%%%%%%%%%%%%%%%%%%%%%%%%%%%%%%
%%%   P   R   O   P   O   S   I   T   I   O   N   %%%%%%%%%%%%
%%%%%%%%%%%%%%%%%%%%%%%%%%%%%%%%%%%%%%%%%%%%%%%%%%%%%%%%%%%%%%

\begin{proposition} \label{prop:right}
The graphon cut norm $\|{\bbW}\|_{\square,1}$ and the induced operator norm $\vertiii{{T_\bbW}}_{2,2}$ satisfy
\begin{align}\label{eqn_prop}
   \|{\bbW}\|_{\square,1} 
      \leq \vertiii{{T_\bbW}}_{2,2} 
              \leq \sqrt{8 \|{\bbW}\|_{\square,1}}    .
\end{align}

\end{proposition}

%%%%%%%%%%%%%%%%%%%%%%%%%%%%%%%%%%%%%%%%%%%%%%%%%%%%%%%%%%%%%%
%%%   M   A   I   N       M   A   T   T   E   R   %%%%%%%%%%%%
%%%%%%%%%%%%%%%%%%%%%%%%%%%%%%%%%%%%%%%%%%%%%%%%%%%%%%%%%%%%%%

%We point out that in Proposition 4 of \cite{ruiz2020graphonsp} the cut norm is denoted as $\|{\bbW}\|_{\square,1} =  \|{\bbW}\|_{\square} $ and the operator norm is denoted as $\vertiii{{T_\bbW}}_{2,2} = \vertiii{{T_\bbW}} $. We change the notation here because we need to introduce other related norms. 

This result is obtained by taking a different view of the integral operator~$T_\bbW$. Indeed, notice that since the kernel~$\bbW \in L^\infty$, $T_\bbW$ can also be considered an $L^\infty \to L^1$ operator with norm
\begin{equation}\label{eqn_pqnorm}
   \vertiii{{T_\bbW}}_{\infty,1}
      := \sup_{\| X \|_\infty \leq 1} 
            \left\| 
               \int_{0}^{1} \bbW(u,v) X(u) \, du 
                  \right\|_1.
\end{equation}
This operator norm is closely related to both the cut norm~\eqref{eqn_cut_norm} and the $2,2$-operator norm, as stated in the following lemmas.

%%%%%%%%%%%%%%%%%%%%%%%%%%%%%%%%%%%%%%%%%%%%%%%%%%%%%%%%%%%%%%
%%%   L   E   M   M   A  %%%%%%%%%%%%%%%%%%%%%%%%%%%%%%%%%%%%%
%%%%%%%%%%%%%%%%%%%%%%%%%%%%%%%%%%%%%%%%%%%%%%%%%%%%%%%%%%%%%%

\begin{lemma}[{\cite[Lemma 8.11]{lovasz2012large}}]\label{lemma_cut_norms}
The cut norm $\|{\bbW}\|_{\square,1}$ [cf. \eqref{eqn_cut_norm}] and the operator norm $\vertiii{{T_\bbW}}_{\infty,1}$ [cf. \eqref{eqn_pqnorm}] satisfy
\begin{align}
   \|{\bbW}\|_{\square,1} 
      \leq \vertiii{{T_\bbW}}_{\infty,1}
            \leq 4 \|{\bbW}\|_{\square,1}  .
\end{align}

\end{lemma}

%%%%%%%%%%%%%%%%%%%%%%%%%%%%%%%%%%%%%%%%%%%%%%%%%%%%%%%%%%%%%%
%%%   L   E   M   M   A  %%%%%%%%%%%%%%%%%%%%%%%%%%%%%%%%%%%%%
%%%%%%%%%%%%%%%%%%%%%%%%%%%%%%%%%%%%%%%%%%%%%%%%%%%%%%%%%%%%%%

\begin{lemma}[{\cite[Lemma E.6]{jansen} with $p=q=2$}]\label{lemma_operator_norms}
The $2,2$-operator norm [cf. \eqref{eqn_22norm}] and the $\infty,1$-operator norm [cf. \eqref{eqn_pqnorm}] satisfy
\begin{align}
        \vertiii{{T_\bbW}}_{\infty,1}
           \leq \vertiii{{T_\bbW}}_{2,2} 
                   %\leq \sqrt{2} \vertiii{{T_\bbW}}_{\infty,1}^{1/2}
                        \leq \sqrt{2 \vertiii{{T_\bbW}}_{\infty,1} } .
\end{align}

\end{lemma}

\begin{remark}
    Note that~\cite[Lemma E.6]{jansen} is written in terms of the ``type-2 cut norm''~\cite[Equation (4.3)]{jansen}:
    \begin{align}\label{eqn_cut_norm_type_2}
       \|\bbW \|_{\square,2} 
          = \sup_{\|f\|_\infty, \|g\|_\infty \leq 1}
               \left | 
                   \int_{0}^{1} \int_0^1 \bbW(u,v) f(u) g(v)\, du dv \right |.
    \end{align}
    It is easy to see that~\eqref{eqn_cut_norm_type_2} is an equivalent definition of the $\infty,1$-operator norm~\eqref{eqn_pqnorm}, i.e., $\|\bbW \|_{\square,2} = \vertiii{{T_\bbW}}_{\infty,1}$~\cite[Remark 4.2]{jansen}.
\end{remark}

%%%%%%%%%%%%%%%%%%%%%%%%%%%%%%%%%%%%%%%%%%%%%%%%%%%%%%%%%%%%%%
%%%   M   A   I   N       M   A   T   T   E   R   %%%%%%%%%%%%
%%%%%%%%%%%%%%%%%%%%%%%%%%%%%%%%%%%%%%%%%%%%%%%%%%%%%%%%%%%%%%

Proposition \ref{prop:right} follows directly from Lemmas \ref{lemma_cut_norms} and~\ref{lemma_operator_norms} as we show next.

\medskip \noindent \textbf{Proof of Proposition \ref{prop:right}.} The lower bound in~\eqref{eqn_prop} is obtained directly from the first inequality in Lemma \ref{lemma_cut_norms} and the first inequality in Lemma \ref{lemma_operator_norms}. For the upper bound, combine the second inequality in Lemma \ref{lemma_cut_norms} and the second inequality in Lemma \ref{lemma_operator_norms} to write
\begin{equation*}
  \vertiii{{T_\bbW}}_{2,2}
     \leq \sqrt{2 \vertiii{{T_\bbW}}_{\infty,1} }
              \leq \sqrt{2 \big(4 \|{\bbW}\|_{\square,1}\big)}.
\end{equation*}
Group terms in the last inequality to conclude the proof. \hfill \qed

We point out that the proofs of Lemmas \ref{lemma_cut_norms} and~\ref{lemma_operator_norms} are not difficult, but for clarity of presentation we prefer to refer the reader to \cite{lovasz2012large, jansen} for their proofs.

\section{Errata in Proposition 4 of \cite{ruiz2020graphonsp}}

After the introduction of Proposition 4 in \cite{ruiz2020graphonsp} we state

\begin{quote}

\medskip

{``This [Proposition 4] is a direct consequence of \cite[Thm. 3.7(a)]{borgs2008convergent} and of the fact that $t(\bbC_2,\bbW)$ is the Hilbert-Schmidt norm of $T_\bbW$, which dominates the $L^2$-induced operator norm.''}

\medskip

\end{quote}

\noindent This argument was appealing to us because it saved the longer argument that we present in Section \ref{sec_proposition} of this note.

However, the argument is incorrect because $t(\bbC_2,\bbW)$ is not, in fact, the Hilbert-Schmidt norm of $T_\bbW$. To see this, recall the definition of the density of homomorphisms from a \emph{simple, undirected} graph $\bbF = (\ccalV,\ccalE)$ onto the graphon $\bbW$:
\begin{equation} \label{eqn:homo_density}
    t(\bbF,\bbW) = \int_{[0,1]^\ccalV}\prod_{(i,j) \in \ccalE} \bbW(u_i,u_j) \prod_{i \in \ccalV} du_i \text{.}
\end{equation}
The simple, undirected version of~$\bbC_2$ has a single edge, i.e., $\ccalE=\{(1,2)\}$. The edge $(2,1)$ cannot be in $\ccalE$ since the motif $\bbF$  must be undirected in~\eqref{eqn:homo_density}. Indeed, note more generally that directed motifs cannot be sampled from $\bbW$. We then conclude that
\begin{align}
    t(\bbC_2,\bbW) &= \int_0^1\int_0^1 \bbW(u,v) du dv
        \text{.}
\end{align}
The above is different from the Hilbert-Schmidt norm of $T_\bbW$, which is defined as
\begin{align*}
  \|T_\bbW\|^2_{\mbox{\scriptsize HS}} 
     = {\int_0^1\int_0^1 |\bbW(u,v)|^2dudv} 
        \text{.}
\end{align*}
%
%Our mistake is glaring in retrospective because it is well known that graphon sequences that converge in the cut metric do not converge in the Hilbert-Schmidt metric \red{\cite{??}}. 

Having acknowledged this mistake, we emphasize that it does not change any of the other results in \cite{ruiz2020graphonsp}. No other proof relies on the incorrect statement that ``$t(\bbC_2,\bbW)$ is the Hilbert-Schmidt norm of $T_\bbW$.'' This is also the only point in the manuscript where we invoke the Hilbert-Schmidt norm. All other results in the paper rely on the convergence of the operator norm  $\vertiii{{T_\bbW}}_{2,2}$, which, as per Proposition \ref{prop:right}, is equivalent to the cut norm.

\subsection{Acknowledgement}

We thank Dr.~Feng Ji for pointing out the mistake in the proof of \cite[Proposition 4]{ruiz2020graphonsp}. Along with collaborators, Dr.~Ji published an extensive correction of our results~\cite{jian2023comments}. These corrections are unnecessary since \cite[Proposition 4]{ruiz2020graphonsp} holds. We simply provided the wrong proof for this result. We also point out that the results in~\cite{jian2023comments} require graph and graphon filters of finite length. This is weaker than the results we provide in  \cite{ruiz2020graphonsp}, which only require filters to have a Lipschitz continuous spectral representation.

\bibliographystyle{IEEEtran}
\bibliography{myIEEEfull,myIEEEabrv,bib-transferability}

\begin{thebibliography}{1}
\providecommand{\url}[1]{#1}
\csname url@samestyle\endcsname
\providecommand{\newblock}{\relax}
\providecommand{\bibinfo}[2]{#2}
\providecommand{\BIBentrySTDinterwordspacing}{\spaceskip=0pt\relax}
\providecommand{\BIBentryALTinterwordstretchfactor}{4}
\providecommand{\BIBentryALTinterwordspacing}{\spaceskip=\fontdimen2\font plus
\BIBentryALTinterwordstretchfactor\fontdimen3\font minus \fontdimen4\font\relax}
\providecommand{\BIBforeignlanguage}[2]{{%
\expandafter\ifx\csname l@#1\endcsname\relax
\typeout{** WARNING: IEEEtran.bst: No hyphenation pattern has been}%
\typeout{** loaded for the language `#1'. Using the pattern for}%
\typeout{** the default language instead.}%
\else
\language=\csname l@#1\endcsname
\fi
#2}}
\providecommand{\BIBdecl}{\relax}
\BIBdecl

\bibitem{ruiz2020graphonsp}
L.~Ruiz, L.~F.~O. Chamon, and A.~Ribeiro, ``Graphon signal processing,'' \emph{IEEE {T}ransactions on {S}ignal {P}rocessing}, vol.~69, pp. 4961--4976, 2021.

\bibitem{jansen}
S.~Janson, ``Graphons, cut norm and distance, couplings and rearrangements,'' \emph{NYJM Monographs}, no.~4, 2013.

\bibitem{lovasz2012large}
L.~Lov{\'a}sz, \emph{Large networks and graph limits}.\hskip 1em plus 0.5em minus 0.4em\relax American Mathematical Society, 2012, vol.~60.

\bibitem{borgs2008convergent}
C.~Borgs, J.~T. Chayes, L.~Lov{\'a}sz, V.~T. S{\'o}s, and K.~Vesztergombi, ``Convergent sequences of dense graphs {I}: Subgraph frequencies, metric properties and testing,'' \emph{Advances in Mathematics}, vol. 219, no.~6, pp. 1801--1851, 2008.

\bibitem{jian2023comments}
X.~Jian, F.~Ji, and W.~P. Tay, ``Comments on '{G}raphon {S}ignal {P}rocessing','' \emph{arXiv:2310.14683 [eess.SP]}, 2023.

\end{thebibliography}

%\section*{Appendix: Proofs of Lemmas} 

%\red{If we choose to include them.}

\end{document}